\documentclass[twocolumn,amsmath,amssymb,aps,pra,superscriptaddress,reprint,notitlepage]{revtex4-1}
\usepackage{pdfsync}
\usepackage{amsmath,amsfonts}
\usepackage{color}
\usepackage{hyperref}
\usepackage{graphicx}
\usepackage{subfigure}
\newcommand{\bra}[1]{\ensuremath{\langle #1|}}
\newcommand{\ket}[1]{\ensuremath{|#1 \rangle}}
\newcommand{\braket}[2]{\ensuremath{\langle #1| #2 \rangle}}
\newcommand{\ketbra}[2]{\ensuremath{| #1 \rangle\hspace{-2pt} \langle #2 |}}
\newcommand{\braoket}[3]{\ensuremath{\langle #1 | #2 | #3 \rangle}}

\newcommand{\eref}[1]{(\ref{#1})}
\newcommand{\fref}[1]{figure \ref{#1}}

\newcommand{\llrr}[1]{\ensuremath{\left( #1\right)}}

\newcommand{\rings}{_\textnormal{R}}
\newcommand{\cells}{_\textnormal{D}}

\begin{document}
\title{Which-way interference within biomimetic unit-cells for efficient energy transfer}
\author{Davide Ferracin}
\affiliation{Quantum Technology Lab, Dipartimento di Fisica, Universit\`a degli Studi di Milano, via Celoria 16, I-20133 Milano, Italy}
\author{Andrea Mattioni}
\affiliation{Institute of Theoretical Physics, Universit\"at Ulm, Albert-Einstein-Allee
11, D-89069 Ulm, Germany}
\author{Stefano Olivares}
\affiliation{Quantum Technology Lab, Dipartimento di Fisica, Universit\`a degli Studi di Milano, via
Celoria 16, I-20133 Milano, Italy}
\author{Felipe Caycedo-Soler}
\affiliation{Institute of Theoretical Physics, Universit\"at Ulm, Albert-Einstein-Allee
11, D-89069 Ulm, Germany}
\author{Dario Tamascelli} 
\affiliation{Quantum Technology Lab, Dipartimento di Fisica, Universit\`a degli Studi di Milano, via
Celoria 16, I-20133 Milano, Italy}
\affiliation{Institute of Theoretical Physics, Universit\"at Ulm, Albert-Einstein-Allee
11, D-89069 Ulm, Germany}

\begin{abstract}
We show that  {\em which-way}  interference  within unit-cells  enhances the propagation along
linear arrays made upon these basic units. As a working example, we address the exciton transfer
through  linear aggregates of ring-like unit cells, the latter resembling the circular structure of
the Light-Harvesting complexes of purple bacteria. After providing an analytic approximate solution
of the eigenvalue problem for such aggregates, we show that the population transferred across the
array is not a monotonic function of the coupling between nearest-neighbor rings, contrary to what
is found from situations where this intra-unit cell interference is not displayed. The
non-monotonicity depends  on an interesting  trade off between the exciton transfer speed and the
amount of energy transferred, which is associated with the rupture of symmetry  among paths within the ring-like cells, due to the inter-ring coupling strength.
\end{abstract}
\maketitle

\section{Introduction} \label{sec:intro}

Recently, the exact nature of the excitonic transfer within the primary steps of photosynthesis has
been under close scrutiny \cite{Scholes_2017RevNat,Romero_2017Rev,CP_HuelgaPlenio}. Numerous
experiments have revealed the presence of persistent oscillatory signals in the spectral response of
different natural light harvesting antennas
\cite{Flemming2005,Engel2007,Lee2007,ColliniWW+10,panit2010,Engel_2018Chem,Scholes_2018NChem},
including the LH2 complex of photosynthetic purple bacteria \cite{Engel_PNAS2012,Hildner2013,
Fidler_2014NComm}. This triggered a large amount of theoretical reasearch aimed to understand the
processes that may underpin the observed  long-lasting coherences
\cite{Ishizaki_2009,ChinPR+13,Christensson_JPCB2012,Jonas_PNAS2012,Womick2011,FelipeCaycedo2012,Kolli2012}.
Even though the biological environments may hinder the relevance of coherent dynamics for
functionality  \cite{Strumpfer_JCP2012,Blau_2018PNAS},  the experiments on these complexes have
prompted a renewed interest on the advantages for  energy/information transfer  provided by quantum dynamics \cite{sension2007,oyala2008,plenio2008,sarovan2010,smyth2012,sangwoo12,Mohseni2008,Lovett_2014PRA,
MdelRey2013,childs02,childs03,christandl05,tamascelli16,bose07}.

Following this strategy, we undertake the study of excitonic energy transfer (EET) in arrays of structures inspired by the geometry
of purple bacteria. The LH2 complex is a circular aggregate composed of eight or nine units, each
with three  bacteriochlorophyll (BChl) molecules, altogether arranged in two concentric rings. Two
of these BChl are responsible for absorption at about 850 nm, and one at about 800 nm, hence referred to as B850 and B800 pigments, respectively.
The biomimetic units we are going to consider in this work will have a simpler geometry and are based on the structure of the B850 ring only. 
The symmetry of the resulting configuration has been recently shown to exhibit {\em which-way} path interference \cite{Tamascelli_2017}. Namely, if a specific pigment is selectively excited, the subsequent propagation along the two available semicircular paths results in a large population in the opposite end of the ring due to the constructive interference of the propagating  wave-packets. In this work, we study the potential use of this interference, aiming to understand which possible ring configurations may benefit from this type of coherent excitation transfer.  Our results show that linear arrays of circular structures without the disorder proper of physiological environments can, indeed, profit from such interference phenomena for efficient energy transfer.

In more detail, we investigate and characterize
EET along linear aggregates of $N$-cycle graphs, or rings,
inspired by the actual structure of the LH2 B850 ring, whose spectral properties are explained in detail in Section II. In Section III, we provide an approximate analytical solution of the eigenvalue problem regarding the linear
aggregates of biomimetic rings. There, we study the dependence of the excitonic transfer rate and the efficiency on the coupling strength
between nearest-neighbor LH2-like rings. We identify, in particular, the optimal values of such
coupling that maximize the transfer efficiency between the end-points of the linear aggregate. Section IV is devoted to a discussion of the results, before drawing our conclusions and perspectives.

\section{Ring-like biomimetic unit cells}

\begin{figure}[t]
    \subfigure[]{\includegraphics[width = 0.5 \columnwidth]{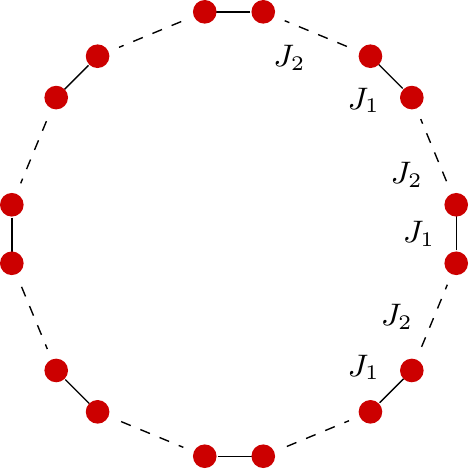}}
    \\[5pt]
    \subfigure[]{\includegraphics[width = 1. \columnwidth]{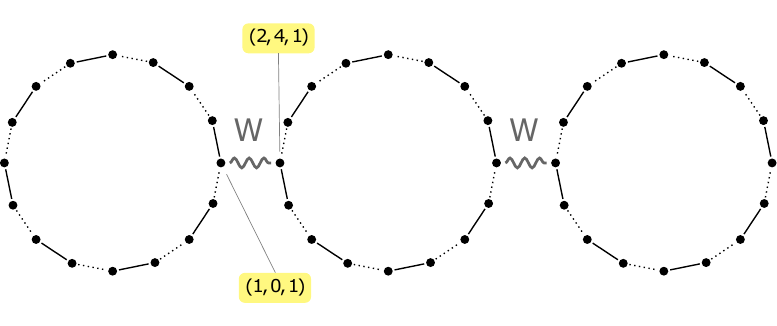}}
    \caption{(a) Geometry of the B850-mimetic ring.  (b) A linear aggregate model with $N\rings=3$ and $N\cells=8$. The pigment-to-pigment coupling   between nearest-neighbor rings is represented by a solid curly line. The figure exemplifies the site-numbering adopted in the paper.}
\label{fig1}
\end{figure}

The schematics of the biomimetic $C_N$ symmetry structure, with $N=8$ cycles is illustrated in
Fig.\ref{fig1}. Each point represents a BChl pigment, that we model as a two level system,
organized in $N\cells$ dimers  $(\text{BChl})_2$. 


We number the dimers with the index $n\in\mathbb{Z}_{N\cells}=\{0,1,\ldots,N\cells-1\}$, and the intra-dimer pigments with $s\in\{1,2\}$. 
We concentrate on the single excitation
manifold spanned by the states
 \[
 \ket{r,n,s}=\ket{0}\otimes\ldots\otimes\underset{(r,n,s)}{\ket{1}}\otimes\ldots\otimes\ket{0},\]
 representing an excitation in the pigment $s\in\{1,2\}$ within the dimer $n\in\{1,\ldots, N\cells\}$
 at ring $r\in\{1,\ldots, N\rings\}$.
The biomimetic unit Hamiltonian 

\begin{equation}\label{full_H}
H = \sum_{r=1}^{N\rings} \ketbra{r}{r}H_{r} + \sum _{r=1}^{N\rings} (\ketbra{r}{r+1}H_{r,r+1}+h.c.)
\end{equation}
is split into the interaction between pigments within a single ring  
\begin{align}
    H_{r}& =
    \sum_{n=0}^{N\cells-1} \sum_{s=1}^2 E_{n,s}\ketbra{n,s}{n,s}+\label{eq:hamiltonian-lh}\\
    &\sum_{n=0}^{N\cells-1} J_1 \llrr{\ketbra{n,1}{n,2}+\ketbra{n,2}{n,1}} \nonumber\\
    & + J_2 \llrr{\ketbra{n+1,1}{n,2}+\ketbra{n,2}{n+1,1}} \nonumber,
\end{align}
where we consider only the interaction between next-neighbor pigments regarding  the same $(J_1)$ or
different $(J_2)$ dimers as shown in \fref{fig1}(a).  
Using the rotational invariance  of the ring Hamiltonian $H_r$ for identical pigments $E_{n,s}=E$, solutions for eigenenergies    $\epsilon_{k,\sigma}$ and eigenvectors  $\ket{k,\sigma}$ of $H_r$ are available \cite{Davidov}
\begin{equation}
    \epsilon_{k,\sigma}=E+(-1)^\sigma\sqrt{J_1^2+J_2^2+2J_1J_2\cos\left (\frac{2\pi
    k}{N\cells}\right )}\label{energies}
\end{equation}
\begin{equation}
    \ket{k,\sigma}=\frac{1}{\sqrt{2}}\bigl[(-1)^\sigma e^{i\eta_k(\beta)}\ket{k,1}+\ket{k,2}\bigr].
    \label{eigenstates}
\end{equation}
where $\eta_k(\beta) =-\arctan\llrr{\frac{\sin\bigl(\frac{2\pi}{N\cells}k\bigr)}{\beta+\cos\bigl(\frac{2\pi}{N\cells}k\bigr)}}$. Here, $\beta=J_1/J_2 >1$ indicates the degree of dimerization of the biomimetic ring and sets a low ($\sigma=1$) and a high ($\sigma=2$) energy manifold. The energies present a pairwise degeneracy for states $\ket{\pm k,\sigma}$ except for $k=0$ and, if $N\cells$ is even, for $k=N\cells/2$. This spectrum, as will be shown here, remains robust for quite large values of the interaction energies $W_{r,n,s,r+1,n',s'}$.


We are interested, however, on the excitation dynamics across a linear array of $N\rings$ rings, which is described by the interaction between pigments of different rings
\begin{align}
    H_{r,r+1}& =\sum_{n,s,n',s'}W_{r,n,s,r+1,n',s'}\ketbra{n,s}{n',s'}.
   \end{align}

Following a similar procedure to the one used for the solutions of \eref{energies} and
\eref{eigenstates},  which is based on rotational invariance of $H_r$, now we make use of the
translational invariance associated with identical rings in order to introduce the states 
$\ket{\rho}=\frac{1}{\sqrt{M}}\sum_{r\in\mathbb{Z}_M} e^{i \frac{2\pi}{M}\rho r}\ket{r}$, with $M=2
N\rings +2$ and $\rho \in \left [-\frac{N\rings}{2},\frac{N\rings}{2} \right ) \cap \mathbb{Z}$.
    Notice that  the antisymmetric combination 
\begin{equation}
    \ket{\rho^-}=\frac{1}{\sqrt{2}}\bigl(\ket{\rho}-\ket{-\rho}\bigr)
    \label{eq_antisymm}
\end{equation}
for $\rho\in\{1,\ldots,\frac{M}{2}-1\}=\{1,\ldots,N\rings\}$, together with $\ket{\rho=0}$ and
$\ket{\rho=\frac{M}{2}}$  are  eigenstates of
$\sum_r (\ketbra{r}{r+1}+\ketbra{r+1}{r})$,  satisfying a vanishing amplitude on the auxiliary rings $r=0$ and $r=\frac{M}2$, as required for the physical aggregate \cite{physchem5b04804}.  

In our model, presented schematically in \fref{fig1}(b), we assume that only the nearest pigments between  adjacent rings are coupled, with an energy $W$, hence 
\begin{equation}
    H_{r,r+1}=
    \begin{cases}
        W\ketbra{0,1}{\frac{N\cells}{2},1} & \mbox{if $N\cells$ is even;}\\
        W\ketbra{0,1}{\frac{N\cells-1}{2},2} & \mbox{if $N\cells$ is odd}.
    \end{cases}
    \label{eq:inter-ring-interaction}
\end{equation}
This leads, for identical rings $H_r=H_0$, to a Hamiltonian that can be written as  $H=\bigoplus_{\rho=1}^{\frac{M}{2}-1} h_\rho$ in terms of  blocks 
\begin{equation}
    h_\rho=H_0+W\cos\llrr{\frac{2\pi\rho}{M}}\hat{V},
    \label{eq:h_rho}
\end{equation}
with
$ \hat{V}=(H_{r,r+1}+H_{r+1,r})/W$.
We will not try to find the exact eigenstates of the system, but we will rather treat the
interaction between the rings $\mathcal{V}=W\cos
\llrr{\frac{2 \pi \rho}{M}}\hat{V}/J_2=\xi_\rho \hat{V}$, where $\xi_\rho$ becomes the perturbation on the single ring Hamiltonian $H_0/J_2$.  Projecting $\mathcal{V}$ onto the degenerate subspace $\{\pm k,\sigma \}$, we obtain
$
    \frac{(-1)^k}{N\cells}\xi_\rho
    \begin{pmatrix}
        1                       & e^{-2 i \eta_k(\beta)} \\
        e^{2 i \eta_k(\beta)} & 1
    \end{pmatrix}$
and calculate the first order correction to the rescaled eigenvalues
\begin{align}
&  \Delta\epsilon_{\rho,k,\sigma,\nu}^{(1)}\label{delta_Es}\\
&=\begin{cases}
    (-1)^k\frac2{N\cells}\delta_{\nu,2}\xi_\rho &\mbox{, $N\cells$ even}\\
    \frac{(-1)^{\sigma+k}}{N\cells}\biggl[\cos\biggl(\frac{\pi k}{N\cells}+\eta_k(\beta)\biggr)+(-1)^\nu\biggr]\xi_\rho  &\mbox{, $N\cells$ odd}
\end{cases}\nonumber
\end{align}
with $\nu\in\{1,2\}$. The corresponding eigenstates, for $k\in\{1,\ldots,\frac{N\cells}2-1\}$ and
$N\cells$ even, are
\begin{equation}
    \ket{\rho,k,\sigma,\nu}=
    \ket{\rho}\otimes\frac{1}{\sqrt{2}}\bigl[(-1)^\nu e^{-2 i
    \eta_k(\beta)}\ket{k,\sigma}+\ket{-k,\sigma}\bigr],
    \label{eq:perturbed-eigenstates-zero-order-even}
\end{equation}
while, for $k\in\{1,\ldots,\frac{N\cells-1}2\}$ and $N\cells$ odd, they read
\begin{align}
    \ket{\rho,k,\sigma,\nu}=
    \ket{\rho}\otimes \frac{1}{\sqrt{2}}\bigl[(-1)^\nu e^{ i \frac{\pi k}{N\cells}- i \eta_k(\beta)}\ket{k,\sigma}+\ket{-k,\sigma}\bigr].
    \label{eq:perturbed-eigenstates-zero-order-odd}
\end{align}


\begin{figure}[t]
    \subfigure[]{
        \includegraphics[width = .48
    \columnwidth]{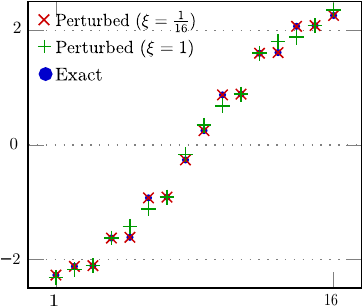}}
    \subfigure[]{
        \includegraphics[width = .48
    \columnwidth]{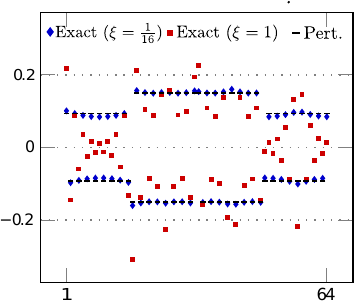}}
    \caption{A comparison between the numerical diagonalization and approximate eigenvalues (a) and
        lowest energy state coefficients (b) for a linear aggregate of
        $N\rings=4$ rings of $N\cells = 8$ dimers for different values of the inter-ring coupling
        strength $\xi= W/J_2$. Each ring is parametrized as in \fref{fig1}.}
\label{fig:exact-perturbed-comparison}
\end{figure}


To settle down an analysis with the prospect of future implementations, we use the couplings
obtained from optical measurements in  the B850 ring \cite{vanGrondelle_1998Biochem}    $J_1=320$
cm$^{-1}$, $J_2=255$ cm$^{-1}$. As shown in figure \ref{fig:exact-perturbed-comparison}(a),
referring to the manifold $\rho=1$, for small $\xi$ the structure of  degenerate doublets ordered in
a high and a low energy submanifold  given by (\ref{energies}) remains, and the result of numerical diagonalization is pretty close to that of the perturbed analysis.  Although the structure of low and high energy submanifolds remains, the pairwise degeneracy is lifted for a few pairs when $\xi=1$ and, interestingly enough, even for this value the perturbative analysis results in a maximum relative error of about   10\%. Hence, concerning  the eigenvalues,  the approximation obtained by the
perturbative analysis is very good, but it results in a worse agreement with the numerically exact
eigenstates, as shown for the amplitudes of the lowest energy state in figure
\ref{fig:exact-perturbed-comparison}(b). However, since the perturbation is proportional to
$\cos \llrr{\frac{2\pi\rho}{M}}$, it acquires its maximum value for the manifold of lowest energy $\rho=1$,  hence, with an approximation to the numerical diagonalization that represents the worst case scenario among all manifolds. Since other manifolds will present even smaller corrections we can safely address the indexes $(k,\sigma)$ as good quantum  numbers for single ring states, even when we consider \emph{all} the interactions of the Hamiltonian $H$.

\section{Characterization of the excitation transfer process}
\newcommand{\muset}{\mathscr{A}}
\newcommand{\prob}{\textnormal{P}}
\newcommand{\kset}{\mathscr{K}}
\newcommand{\ptset}{\mathscr{B}}

We will now use a collective index $\{\mu\}=\{(\rho,k,\sigma,\nu)\}=\{1,\cdots, 2N\cells N\rings\}$ to ease the notation labeling  the eigenstates of $H$,  $\ket{e_\mu}$, which are indexed  in the ascending order of their respective energies $\epsilon_\mu$.  

With these eigenstates and eigenvalues, we can approach the study of the evolution in time of the excitation as it is transferred along the chain from a given initial state $\ket{\psi(t=0)}=\ket{\psi_0}$, after a time $t$, $  \ket{\psi(t)}=\exp(-i Ht)\ket{\psi_0}$. 

 The probability that the excitation is found in the $r$-th  ring after $t$ is 
\begin{align}
    \prob_r&(t)=
    \sum_{n\in N\cells,s} \left |{\bra{r,n,s}\exp(-i Ht)\ket{\psi_0}}
    \right |^2\nonumber\\
& = \sum_{\mu,\mu'}
    e^{i(\epsilon_{\mu'}-\epsilon_{\mu}) t}
    \bra{e_{\mu'}}\pi_r
    \ket{e_{\mu}}\braket{e_\mu}{\psi_0}\braket{\psi_0}{e_{\mu'}}
    \label{eqPt}
\end{align}
where $\pi_r=\ket{r}\bra{r}\otimes\sum_{n \in N\cells ,s} \ket{n,s}\bra{n,s}$. Due to the identification of $\ket{k,\sigma}$ as appropriate quantum numbers for the rings in the chain, this projector can also be written as $\pi_r=\ket{r}\bra{r}\otimes\sum_{k'',\sigma''}\ketbra{k'',\sigma''}{k'',\sigma''}$ and the approximation $\bra{e_{\mu'}}\pi_r
    \ket{e_{\mu}}\approx \delta_{k',k}\delta_{\sigma',\sigma}\delta_{\nu',\nu}\braket{\rho'}{r}\braket{r}{\rho}$ follows. This results in
\begin{align}
    \prob_r&(t) \approx \sum_{\rho,\rho',k,\sigma,\nu}e^{i(\epsilon_{\rho',k,\sigma,\nu}-\epsilon_{\rho,k,\sigma,\nu }) t}\braket{\rho'}{r}\braket{r}{\rho}\nonumber\\
&\times\braket{\rho,k,\sigma,\nu}{\psi_0}\braket{\psi_0}{\rho',k,\sigma,\nu}
\label{Pr_approx}
\end{align}
which together with the solutions (\ref{delta_Es}-\ref{eq:perturbed-eigenstates-zero-order-even}), represents the perturbative solution for the dynamics of ring populations.
 
\begin{figure}   
    \includegraphics[width=1.\columnwidth]{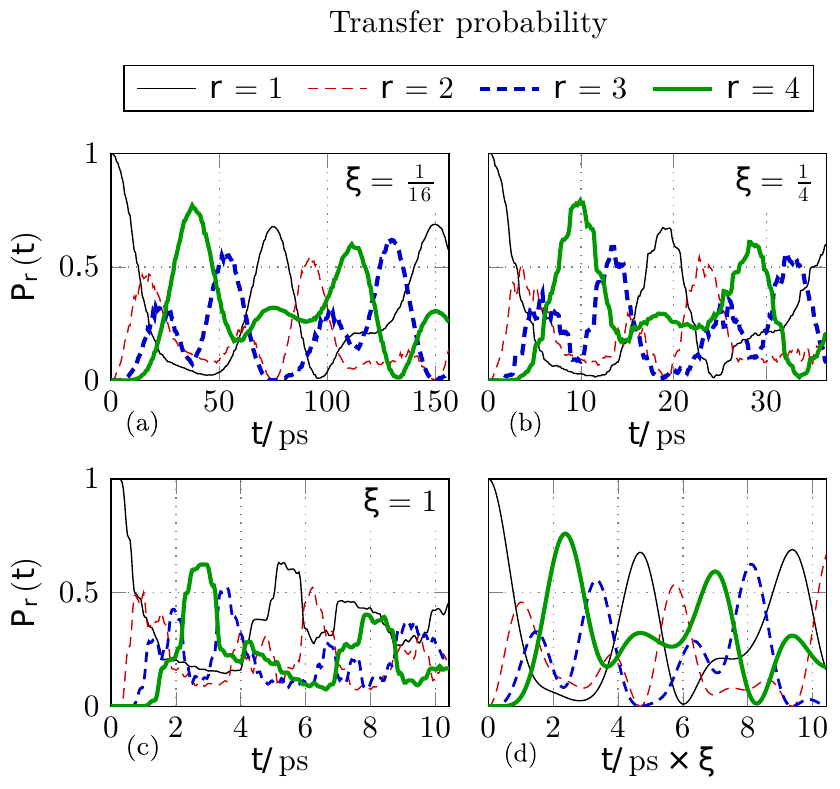}
    \caption{Time evolution of populations in an aggregate of 4 rings, when initially the excitation
        is localized on the leftmost site of the first ring ($\ket{\psi(0)} = \ket{1,4,1}$) for
    different values of the inter-ring coupling $\xi$. Panel (d) shows the result of the perturbative analysis presented as a function of  the rescaled time $\xi t$.}
\label{fig:transfer-localised-NR4-NC8}
\end{figure}

In figure \ref{fig:transfer-localised-NR4-NC8} we present the time evolution of the populations in a linear aggregate of four rings, for different values of $\xi$ and using the same parameters we chose for the eigenvalues and eigenstates of figure \ref{fig:exact-perturbed-comparison}.
First, notice that the solution from numerical diagonalization for $\xi$ equal to $\frac1{16}$ or
$\frac14$  presented in panels (a) and (b), is very similar to the perturbative solution which is
shown in panel (d) for $\xi=\frac1{16}$. Second, note that in panels (a-c) care has been taken to adjust the range of the time axis in a way that it is inversely proportional to the value of $\xi$, which is made explicit in the presentation of the perturbative solution, with the scaled  time $t \xi$. Observe that equation (\ref{Pr_approx}) depends on time solely through the argument  $t(\epsilon_{\rho,k,\sigma,\nu}-\epsilon_{\rho',k,\sigma,\nu })$, which can be finite if the translational invariance across the aggregate of rings is lifted, accomplished by the coupling $W$ between them. 
Following the solution to first order perturbation on $\xi$ addressing the degenerate manifold $\{\ket{\pm k,\sigma}\}$, we find that $\prob_{N\rings}(t)\approx\prob_{N\rings}^{(1)}(t)= f(\alpha \xi t)$ where  $\alpha=2/N\cells$ and $f$ is a non-negative smooth function (see Appendix for details). It follows then, that as long as $\xi$ remains small, the time-scale of transfer is given by  $\simeq 1/\xi$ as obtained in figure \ref{fig:transfer-localised-NR4-NC8}. In this regime, therefore, the details of the ring  Hamiltonian, e.g. the dimerization degree $\beta=J_1/J_2$, are not important to determine the properties of the transfer along the aggregate. 

\begin{figure}    \includegraphics[width=1.\columnwidth]{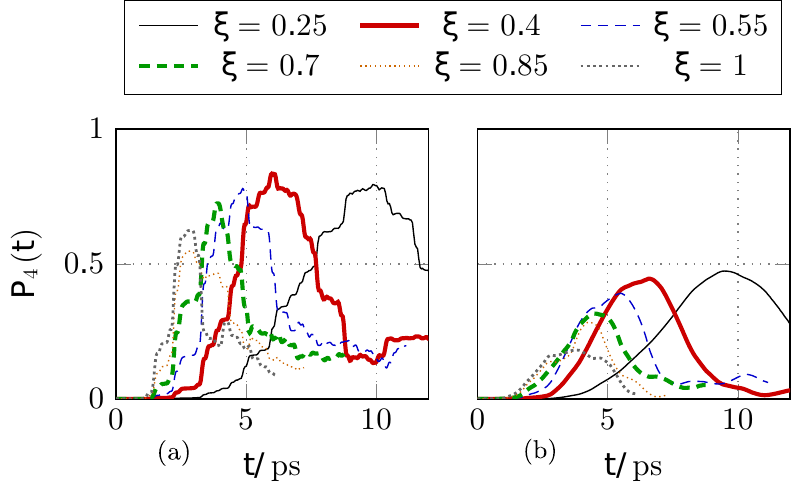}
\includegraphics[width=1.\columnwidth]{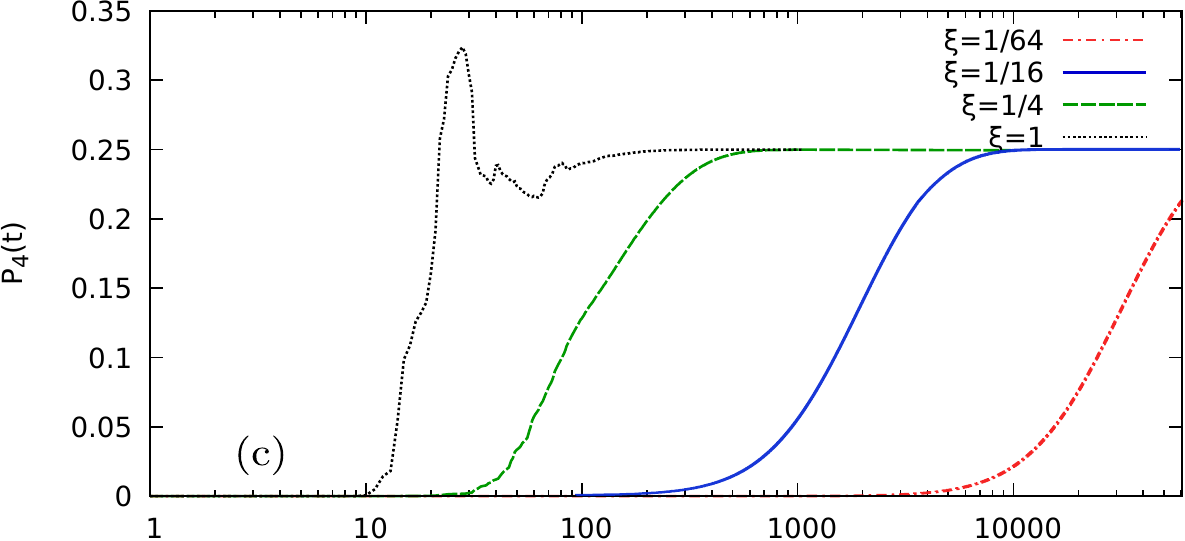}
    \caption{Top panels: the transfer efficiency as a function of time for different values of the rescaled
        inter-ring coupling $\xi=W/J_2$. The rings are parametrized as in the previous figures.
        (a) localized initial condition
        $\ket{\psi(0)} = \ket{1,4,1}$; (b) delocalized initial condition $\ket{\psi(0)}=
        \ket{r=1,k=1,\sigma=1}$.  (c) the effect of dephasing: the same quantity plotted in
        panels (a) and (b) for the same dephasing rate $\Gamma = 0.05 J_2$ and different values of
    the inter-ring coupling $W$ (in units of $J_2$).}
\label{fig:varcoupling}
\end{figure}

Notice in figure \ref{fig:transfer-localised-NR4-NC8}(c) and in more detail in figure
\ref{fig:varcoupling}(a), that  the height of the first peak in the fourth ring becomes smaller  for greater values of $\xi$.  In order to explain this, we need to take into account also second order corrections in $\xi$. As will be shown next, while at first order the perturbation only lifts the degeneracies within the manifolds $\{\ket{\pm k,\sigma}\}$, higher order contributions mix states from non-degenerate manifolds,
introducing in \eqref{eqPt} an additional dependence on $\xi$ through the amplitudes of the
perturbative eigenstates $\ket{e_{\mu}}$.  Using definition \eqref{eq_antisymm}, we can write the
probability to find the excitation in the last ring, in general, as
\begin{align}
\prob_{N\rings}(t)&=\sum_{\rho,\rho'}(-1)^{\rho+\rho'}|\braket{1}{\rho}|^2 |\braket{1}{\rho'}|^2 \nonumber \\ 
 & \times
\sum_{\kappa,\kappa'}
e^{i(\epsilon_{\rho\kappa}-\epsilon_{\rho'\kappa'})t}
\braket{\phi}{\kappa_{\rho}}
\braket{\kappa_{\rho}}{\kappa'_{\rho'}}
\braket{\kappa'_{\rho'}}{\phi},
\label{eq:pnr0}
\end{align}
where $\ket{\phi}$ is the initial state of the first ring, and we denote by $\kappa$ the $\rho$-independent part of the generalized index $\mu=(\rho,\kappa)$ introduced in \eqref{eqPt}.
The states $\ket{\kappa_{\rho}}$ are eigenstates of \eqref{eq:h_rho} with eigenvalues $\epsilon_{\rho\kappa}$.
The subscript $\rho$ indicates that the amplitudes may depend explicitly on the perturbation $\xi_{\rho}$, and therefore  $\braket{\kappa_{\rho}}{\kappa'_{\rho'}}=\delta_{\kappa,\kappa'}$ only when $\rho=\rho'$ or when the amplitudes do not depend on $\rho$, as happens at first order in $\xi$.
We can explicitly see this additional $\xi$-dependence by considering the effect of the inter-ring coupling on two non-degenerate subspaces $\{\ket{\pm k_1,\sigma} \}$ and $\{\ket{\pm k_2,\sigma} \}$ such that $\braoket{k_1,\sigma}{\hat{V}}{k_2,\sigma}\neq 0$ and solving the dynamics in this 4-dimensional subspace only.
For this reduced subspace, we obtain the following expression
\begin{equation}
\prob_{N\rings}(t)\approx \alpha \left( f(\alpha \xi t) + \frac{\alpha^2 \xi^2}{2 \delta^2}f''(\alpha \xi t) \right)
\label{prob_2nd}
\end{equation}
valid to lowest non-zero order in $\xi$, where $\delta=(\epsilon_{k_1,\sigma}-\epsilon_{k_2,\sigma})/J_2$ (see Appendix for details).
To first order in $\xi$,  $\prob_{N\rings}$ reaches a maximum when the first derivative of $f$ is zero and the second derivative $f''$ is negative. From \eqref{prob_2nd} we see that the inter-ring coupling has two effects: on the one hand it compresses the transfer time-scale through the argument $\alpha \xi t$  of $f$, while on the other, it reduces the amplitude of $\prob_{N\rings}$ by an amount proportional to $f''$. 
Notice that the difference in energy $\delta$  depends on the details of the ring, as provided by equation \eqref{energies}, so it follows that the amplitude of the transfer can be devised upon the details of the ring interactions.
 It turns  out then, that the  speed (depending on the coupling strength between rings) and
 amplitude (depending on the geometrical details and couplings within each ring) of the transferred
 population maximize to a ``sweet spot'', as expected from concurrent quantities.

To better resolve this optimal value, figure \ref{fig:varcoupling}(a) shows that the
amplitude of the transferred population starts to decrease appreciably when $\xi$ is between ${0.4}$ and ${0.5}$, while the time required to reach
the maximum population in the fourth ring decreases from about 4 to 3 ps. A further increase to  $\xi=0.7$  reduces this time to 2 ps  at the expense of reducing about 15\% the amplitude of the last ring population. 
Changes of the ring geometry/interactions may suppress this reduction, but a specific parameter set can be better implemented if additional dissipation processes that may avoid the successful transfer to the end of the chain are incorporated.

%
%

 In order to shed  light on the phenomenon engaged in the variable amount of population transferred
 to the last ring, in figure \ref{fig:varcoupling} we compare the population transferred to the last
 ring when the first ring is initialized on a single pigment  (a),   or delocalised over the full
 ring  (b). As it can be observed, the former initialization leads to a much higher transfer
 efficiency than the latter. Based on \eref{eq:pnr0}, the population in the last ring  for a
 localized initial state  $\prob_{N\rings}^{\text{\scriptsize loc}}(t)$ doubles that  resulting from
 a delocalized initial state $\prob_{N\rings}^{\text{\scriptsize del}}(t)$, associated with the localized $\ket{\phi}=\ket{\frac {N\cells}{2},1} $ and delocalized $\ket{\phi}=\ket{k=1,\sigma=1} $ states, respectively. The relation $\prob_{N\rings}^{\text{\scriptsize loc}}(t)=2\prob_{N\rings}^{\text{\scriptsize del}}(t)$ holds only if the  initial state is localized either at the coupling site $\ket{n=0,s=1}$ or at its diametrically opposite pigment $\ket{n=\frac{N\cells}{2},s=1}$. This result provides a physical picture where the localized initial state propagates in two wavefronts along the branches of the ring, which  interfere constructively at the site coupled to the neighboring ring, maximizing population transfer. Notice that the choice of a localized initial condition  from any other pigments than those lying across the linear aggregate axis, leads to a transfer probability equal to $\prob_{N\rings}^{\text{\scriptsize loc}}(t)/2$, which is reduced because such initial condition disrupts this quantum interference at the coupling site in the first ring.

However, notice that  the effect shall be attributed to both the ``which-way'' propagation within each ring-like unit-cell, and the particular form of the inter-ring coupling specified in \eqref{eq:inter-ring-interaction}. This latter allows  initialization of intermediate rings at pigments along the linear aggregate axis, which enable the constructive interference to happen in these intermediate stages, even though a delocalized initial conidtion is chosen. Thereby, the reduction $\prob_{N\rings}^{\text{\scriptsize del}}(t)=\frac 12\prob_{N\rings}^{\text{\scriptsize loc}}(t)$ can be attributed to the internal dynamics of the first ring, while an additional reduction should arise if the interference in the intermediate rings is somehow hindered.

To illustrate this point, we calculate the dynamics of the linear aggregate with  dephasing between
pigments. In short, we evolve the density operator $\rho$ according to $\partial
_t\rho=-i[H,\rho]-\Gamma
\sum_{r,n,s}(\rho\ket{r,n,s}\bra{r,n,s}+\ket{r,n,s}\bra{r,n,s}\rho-2\ket{r,n,s}\bra{r,n,s}\rho\ket{r,n,s}\bra{r,n,s})$,
where $\Gamma$ describes the dephasing rate between the ground and the excited state of pigment
$r,n,s$. The results in figure \ref{fig:varcoupling}(c), show that for $\xi=1$ and $2\pi W> \Gamma$,
the oscillatory dynamics results in a maximum population at the last ring ($\approx 0.35$) which is
considerably smaller than the analogue quantity in the fully coherent case ($\approx 0.6$, see  figure
\ref{fig:varcoupling}(a)). Thereby, decoherence within pigments decreases the peak population in the last ring, which underlines the importance of constructive interference within each ring to accomplish efficient transfer.

\section{Conclusion and outlook}

Motivated by the apparent symmetry of natural harvesting structures, we investigated the effect of short range interactions and dynamics within ring-like unit cells, on the long-range  propagation of excitations in linear aggregated of these basic units.  Under the assumption that we can define these unit cells, hence, in a scenario where intra-ring interactions are stronger than inter-ring couplings, we studied the situation in which only the closest pigments of nearest-neighbor rings in the aggregate interact with each other.  We obtained analytic approximate solution of the eigenvalue problem for the system Hamiltonian by means
of a perturbative approach on the inter-ring coupling strength, which enabled us to illustrate that inter-ring coupling is the only quantity that sets the time-scale for coherent population exchange between rings, while the geometry of the unit cells may result in a beneficial constructive interference of wavepackets in the regions where the interaction among unit-cells is strongest. 

Although the magnitude of disorder and the non-local nature of inter-LH2 coupling in biological relevant membranes will hinder these effects from having a functional relevance in photosynthesys, implementations in synthetic systems, such as carbon-linked porphyrin nanorings \cite{OSullivan2011,Kondratuk2014}, could be more promising. These fully $\pi$-conjugated complexes have recently been shown to support
robust quantum interference and a high degree of exciton delocalization \cite{Richert2017,Rickhaus2017,Yong2015,Parkinson2014JPCL},
which are prerequisites for implementation of technologies that utilize the effects that we address in this work.\\
Even though the specific conditions to observe the advantages of constructive interference in unit-cells might be challenging at present, with this work, it is also our desire to motivate the investigation of quantum mechanical dynamics within unit-cells for optimization of transfer at larger scales, as expressed with the Multichromophoric F\"orster Theory formalized some years ago by  Silbey and coworkers \cite{SIlbey_2004PRL}.

\section*{Acknowledgements}
We thank Susana F. Huelga and Martin B. Plenio for many useful discussions. This work was
supported by the ERC Synergy grant BioQ and the IQST. This publication was made possible through the
support of a grant from the John Templeton Foundation.
\section*{Appendix}

\subsection{Perturbative expression of the transfer probability}
The starting point for the calculation of the transfer probability $\prob_{N\rings}(t)$ is \eqref{eq:pnr0}.
The orthogonality condition $\braket{\kappa_{\rho}}{\kappa'_{\rho'}}=\delta_{\kappa,\kappa'}$  valid for $ \rho=\rho'$ or for $\rho$-independent amplitudes (as it occurs at first order in $\xi$) renders the calculation straightforward for the first term in \ref{prob_2nd} and  yields the result
$\prob_{N\rings}(t)\approx f(\alpha\xi t)$, with
\begin{equation}
f(x)=\left| \sum_{\rho} (-1)^{\rho}|\braket{1}{\rho}|^2 e^{i x \cos( \frac{2\pi\rho}{M}) } \right|^2.
\end{equation}
 The calculation can be repeated considering an initially localized state on the first ring 
\begin{align}
\prob_{N\rings}^{\text{\scriptsize loc}}(t)
&=\left| \sum_{\rho} (-1)^{\rho}|\braket{1}{\rho}|^2 e^{i\frac{2}{N\cells}\xi_{\rho} t} \right|^2
\label{eq:ploc} \\
\end{align}
or delocalized in a single ring eigenstate $\ket{k,\sigma}$, which results in the relation
\begin{align}
\prob_{N\rings}^{\text{\scriptsize del}}(t)
&=\frac{1}{2}\prob_{N\rings}^{\text{\scriptsize loc}}(t)
\label{eq:pdel} 
\end{align}

The calculations are more involved when we consider second order mixing between non degenerate manifolds.
In this case we compute the dynamics only within a 4-dimensional subspace given by two degenerate manifolds $\{\ket{\pm k_1,\sigma}\}$ and $\{\ket{\pm k_2,\sigma}\}$.
Projecting \eqref{eq:h_rho} on this subspace, one sees that only two of the four first order eigenstates couple, namely $\{\ket{i}=\braket{\rho }{ \rho,k_i,\sigma,2}\}_{i=1,2}$ (cfr.\eqref{eq:perturbed-eigenstates-zero-order-even} \eqref{eq:perturbed-eigenstates-zero-order-odd} in main text).
These result in the $\kappa_{\rho}$-states
\begin{align}
\ket{+_{\rho}} &= e^{i(\eta_{k_1}-\eta_{k_2})}\cos\theta_{\rho}\ket{1}+\sin\theta_{\rho}\ket{2} \label{eq:kappaplus}\\ \ket{-_{\rho}} &= -e^{i(\eta_{k_1}-\eta_{k_2})}\sin\theta_{\rho}\ket{1}+\cos\theta_{\rho}\ket{2} \label{eq:kappaminus}
\end{align}
with eigenvalues $\epsilon_{\rho,\pm}=\alpha\xi_{\rho}\pm\frac{1}{2}\sqrt{\delta^2+4\alpha^2\xi_{\rho}^2}$,
where the mixing angle is given by $\theta_{\rho}=\frac{1}{2}\arctan
\llrr{\frac{2\alpha\xi_{\rho}}{\delta}}$.
Plugging \eqref{eq:kappaplus} and \eqref{eq:kappaminus} into \eqref{eq:pnr0} gives rise to some terms oscillating at a frequency $O(\delta)$, which is larger than the end to end transfer frequency $O(\xi)$. Therefore we perform a rotating wave approximation by retaining only the terms that oscillate with a frequency of order $\xi$.
At this point, expanding the mixing angles to second order in $\xi$, one obtains the final expression 
\eqref{prob_2nd}, showing the explicit dependence of the probability on the last ring on the perturbative parameter.

%

\end{document}